\documentclass[10pt, conference]{IEEEtran}

\usepackage[table,xcdraw]{xcolor}
\usepackage[hidelinks]{hyperref}
\usepackage{color,soul}
\usepackage{mathtools}
\usepackage{amssymb}
\usepackage{algorithm,algpseudocode}
\usepackage[binary-units,per-mode=symbol,range-phrase=--]{siunitx}
\usepackage{flexisym}
\usepackage[export]{adjustbox}
\usepackage{footnote}
\usepackage{makecell}
\usepackage{comment}
\usepackage{pgfplots,filecontents}
\usepackage{pgfplotstable}
\usepackage{tikz}
\usepackage{subcaption}
\usepackage{graphicx}
\usepackage[inline]{enumitem}
\usepackage{multirow}
\usepackage{diagbox}
\usepackage{url}
\setcellgapes{1pt}
\begin{document}
\title{Scheduling of Distributed Applications on the Computing Continuum: A Survey}

\author{\IEEEauthorblockN{ 
Narges Mehran$^{\star}$, Dragi Kimovski$^{\star}$, Hermann Hellwagner$^{\star}$,
Dumitru Roman$^{\ddagger\mathsection}$, Ahmet Soylu$^{\mathsection}$, 
Radu Prodan$^{\star}$}
\IEEEauthorblockN{$^{\star}$Alpen-Adria-Universit{\"a}t Klagenfurt, {Austria} \\
$^{\ddagger}$SINTEF AS, Oslo, {Norway}\\$^{\mathsection}$OsloMet - Oslo Metropolitan University, Oslo, {Norway}\\
Email: {narges.mehran,dragi.kimovski,hermann.hellwagner,radu.prodan}@aau.at\\
Email: {dumitru.roman,ahmet.soylu}@oslomet.no}
}

\maketitle

\begin{abstract}
The demand for distributed applications has significantly increased over the past decade, with improvements in machine learning techniques fueling this growth. These applications predominantly utilize Cloud data centers for high-performance computing and Fog and Edge devices for low-latency communication for small-size machine learning model training and inference. The challenge of executing applications with different requirements on heterogeneous devices requires effective methods for solving NP-hard resource allocation and application scheduling problems. The state-of-the-art techniques primarily investigate conflicting objectives, such as the completion time, energy consumption, and economic cost of application execution on the Cloud, Fog, and Edge computing infrastructure. Therefore, in this work, we review these research works considering their objectives, methods, and evaluation tools. Based on the review, we provide a discussion on 
the scheduling methods in the Computing Continuum. 
\end{abstract}


\textcolor{red}{\scriptsize 2023 ACM/IEEE.  Personal use of this material is permitted.  Permission from ACM/IEEE must be obtained for all other uses, in any current or future media, including reprinting/republishing this material for advertising or promotional purposes, creating new collective works, for resale or redistribution to servers or lists, or reuse of any copyrighted component of this work in other works.}

\section{Introduction}
Distributed processing applications~\cite{de2018distributed} encompass gathering, storing, and analyzing massive amounts of data, gradually overwhelming centralized \emph{Cloud} data centers. On the other hand, \emph{Fog and Edge computing} technologies extend the Cloud with computation, storage, and network services in the proximity of the Internet of Things (IoT) devices~\cite{mehran2022c3match}. Consequently, media, healthcare, and manufacturing companies, to name a few, can move their latency-sensitive applications from the Cloud data centers to utilize the Fog and Edge devices near data producers and consumers.

As alternatives to the Cloud, infrastructures such as Exoscale Edge~\cite{exoscaleedge}, Google Distributed Cloud~\cite{gdce}, along with local services provisioned by mobile devices~\cite{kimovski2021mobility}, consolidate the Fog and Edge with computing and data management services. The unification of Cloud, Fog, and Edge infrastructures (aka.~\emph{Computing Continuum}) aims to provide a seamless aggregation of distributed resources of all kinds to support the emerging dataflow processing applications~\cite{balouek2019towards}.

Over the past few years, academia and industry investigated the challenges in distributed application execution and scheduling on the Computing Continuum~\cite{kochovski2019architecture}. This work reviews the state-of-the-art methods for scheduling distributed applications on the Computing Continuum provisioned by different resource providers. 

Most scheduling and placement methods prioritize minimizing application completion times, which is paramount in meeting users' service level agreement requirements. Fast completion times are a fundamental objective that the methods address to ensure user satisfaction. Additionally, the timely delivery of services remains a central concern for application scheduling. Moreover, energy efficiency is a critical factor that resource allocation and scheduling methods must consider. To underscore the magnitude of this issue, Google, for instance, approximately consumes \SI{15.5}{\tera\watt\hour} of electricity annually, with a significant portion being attributed to its data centers\footnote{\url{https://www.gstatic.com/gumdrop/sustainability/247-carbon-free-energy.pdf}}. 
In addition, economic cost represents another critical parameter that resource providers prioritize for optimizing infrastructure maintenance expenses, ultimately benefiting their consumers. Hence, after reviewing the research works, we categorize them based on the non-functional objectives, such as the completion time, energy consumption, and economic cost.

The rest of the paper is structured as follows. We survey the literature mainly addressing completion time in Section~\ref{time}. Section~\ref{energy} describes the application scheduling and resource allocation methods with the main objective of minimizing energy consumption. 
Section~\ref{cost} categorizes the cost-optimizing methods, followed by the discussion in Section~\ref{discussion} and the conclusion in Section~\ref{conclusion}.

\section{Completion time}\label{time}
In this section, we begin by discussing the most prominent methods in the literature, exploring the time of the application execution.  
Samani et al.~\cite{samani2023incremental} proposed a multilayer resource partitioning method to minimize resource wastage by utilizing the total capacity of heterogeneous Fog devices. This method places each service on a Fog device in each partition by minimizing the completion time of an application. Moreover, this method aims to maximize deadline satisfaction in the Fog with a high arrival rate of requests.

Skarlat et al.~\cite{Skarlat2017QoS-aware} explored the Fog service placement problem
and modeled it as an integer programming model by optimizing the IoT application's completion time on Fog and Cloud resources. Therefore, in this method, an orchestrator allocates the resources of the Fog data centers to users' applications and proactively collects periodic updates to satisfy the QoS metrics, such as the deadline for the application's completion and processing times.

Veith et al.~\cite{da2018latency} proposed a scheduling strategy called response time rate with region patterns, which uses a greedy strategy to identify the resources that minimize the service completion time on the Cloud and Fog. This method decomposes the application based on dataflow patterns such as a fork or a join and distributes it to the Cloud and Fog computing infrastructures.

Veith et al.~\cite{da2019monte} applied reinforcement learning and the Monte-Carlo tree search algorithm to reschedule application operators during runtime. This method formulates the operator assignment problem as a Markov decision process to minimize the application's completion time. Afterward, the learning algorithm determines the reconfiguration of the resource allocation to the operators.

Xia et al.~\cite{xia2018combining} proposed four combined scheduling heuristics, such as Fog devices and application components ordering, optimizing the weighted average completion time for a large set of heterogeneous applications constrained by a group of non-functional parameters, including processing and storage capabilities.

Souza et al.~\cite{souza2018towards} applied Min-Max heuristics to optimize the service completion time. Best-fit with the queue method performs service atomization to represent parallel and sequential service execution strategies. It analyzes the service request and the expected availability time for Fog devices. The BQ method minimizes the service's transmission and processing times while maximizing the priority of the Edge devices for the microservices waiting in a processing queue. 

Sun et al.~\cite{Sun2018} investigated a two-level scheduling model categorizing Fog devices in clusters based on their locations. Afterward, the scheduling method decides on the appropriate Fog cluster and the proper device within the given cluster. The weighted multi-objective optimization method reduces the service completion time and maximizes the reliability of application execution.

Elgamal et al.~\cite{elgamal2018droplet} proposed an algorithm to partition operations of an IoT application on Edge and Cloud devices to minimize operations' completion times. This method includes data processing, queuing, and the tradeoff between processing and transmission when placing operators on heterogeneous devices.

Khare et al.~\cite{khare2019linearize} presented an optimization-based scheduling method to minimize the application's completion time. This method decomposes an application graph into several linear paths from the application's source to the sink. Afterward, it applies an execution-time prediction model for co-located linear dependencies to minimize the completion times of all paths in the application's graph. 

Ravindra et al.~\cite{ravindra2017mathbb} proposed an open-source orchestration platform for application execution on the Cloud, Fog, and Edge devices. This platform supports streams, micro-batches, and various data processing frameworks like Apache Edgent, Storm, and Spark. The modeled orchestrator supports the migration of application components if the resource reaches an over-utilized state by executing a high workload of operators and application components.

Renart et al.~\cite{renart2019distributed} extended a data analytics software framework called R-Pulsar to solve the placement problem of IoT operators on the Edge and Cloud devices. This method reduces data transfer rate, completion time, and economic cost (based on AWS and Microsoft pricing models). It divides the IoT application graph into paths consisting of sequences of operators. Afterward, the longest path defines the completion time, including data transmission times between operators over the path.

\textit{Summary.} The works investigate a single~\cite{da2018latency,da2019monte}, weighted set of objectives~\cite{elgamal2018droplet,Skarlat2017QoS-aware}, multi-objective~\cite{Sun2018,ravindra2017mathbb,renart2019distributed,souza2018towards}, or predictive~\cite{khare2019linearize} optimization models. They do not minimize the cost and energy consumption of an application's execution on the heterogeneous Computing Continuum. 

\section{Energy consumption}\label{energy}
A growing body of literature has recently described various methods to curb energy consumption.  
Kansal et al.~\cite{kansal2010virtual} proposed a linear regression-based method to reduce the energy consumption of virtual machines. The method considers static and dynamic power consumption of a device's processor, memory, and storage. Therefore, it monitors the resource utilization of virtual machines with the help of a virtualization monitoring tool, retrains the regression-based model, and estimates the energy consumption.

Al Faruque et al.~\cite{al2015energy} proposed a decentralized optimization-based scheme in which distributed controllers provide the energy management service in a Fog computing infrastructure. This method manages the energy consumption of residential buildings and all devices, such as sensors and actuators, along with computing and network devices. The control-as-a-service architecture module manages the devices to decrease energy consumption, cost, and time for their product to be available in the market. 

Shojafar et al.~\cite{shojafar2016energy} proposed an energy-adaptive 
resource scheduler to maximize computational efficiency while maintaining the user's satisfaction. The main goal of this method was to minimize energy consumption while executing the application in the Fog infrastructure. This scheduler consolidates the resources by managing the application execution without reconfiguring virtual machines. 

Pooranian et al.~\cite{pooranian2017novel} designed a penalty-aware heuristic algorithm to minimize processing and transmission energy in the Fog and preserve load balancing among the data centers. Their penalty-aware algorithm assigns a negative score to a Fog data center 
consuming more energy and avoids allocating its resources to applications during the subsequent runtime iteration.

You et al.~\cite{you2018asynchronous} proposed an energy-efficient asynchronous mobile edge computation offloading method. This method solves the optimization-based computation offloading problem to obtain 
a decision between local processing on a mobile device and offloading the computation to the Edge infrastructure.

Reddy et al.~\cite{reddy2020genetic} proposed a genetic-based model to minimize the energy consumption of the Fog devices. This model reduces the number of service requests failing to perform their operations while shortening the overall execution time. This method applies a reinforcement learning model to optimize the energy consumption of Fog devices while executing the applications.

Menouer~\cite{menouer2021kcss} presented a scheduling model based on preference, similar to the Technique for Order Preference by Similarity to Ideal Solution (TOPSIS)~\cite{hwang2012multiple} that optimizes multiple criteria, including resource utilization and energy consumption. A decision-making algorithm reduces the processing, memory, storage resource utilization, and power consumption of resources, aimed at satisfying the user's and resource provider's requirements while balancing them for executing every application.

\textit{Summary.} These works optimize either a single constrained objective~\cite{kansal2010virtual,pooranian2017novel,shojafar2016energy,you2018asynchronous} or bi-objective~\cite{menouer2021kcss} model and do not focus on minimizing cost along with the energy consumption of an application's execution on the heterogeneous Computing Continuum.

\section{Economic cost}\label{cost}
Recently, considerable ongoing research and effort has been dedicated to the economic cost aspect within the Computing Continuum.  
Aazam et al.~\cite{aazam2015fog} proposed a method to manage the Fog devices as a heterogeneous pool for data processing with unpredictable resource demands. Therefore, the resource provider considers the application requirements, such as the type and price of the service, to reduce the economic cost of service execution in the Fog while encouraging the application providers to use this infrastructure.

Zhao et al.~\cite{zhao2016pricing} proposed a price-based competition between the heterogeneous Edge and Cloud resources, while this method balances all the providers' profits. Their proposed method maximizes both providers' profit and users' satisfaction. The Edge resources are more profitable when the cost is a user's essential metric.

De Maio and Kimovski~\cite{demaio-kimovski2020} investigated a parallel task decomposition method that could potentially improve the Fog's completion time, reliability, and monetary cost of extreme data workflows. They evaluated the Pareto-based method with four use-case workflows from the field of natural sciences through the Spark-enabled mobile Edge offloading Monte-Carlo simulator 
and extended FogTorchPI~\cite{Brogi2017}. Moreover, it categorized the workflow's tasks based on their dependencies and devised the independent tasks' offloading to the Fog infrastructure.

Sharghivand et al.~\cite{sharghivand2020edge} proposed a double-sided matching scheduler for data analytics tasks at the Edge based on the device and application preferences toward lower completion time and cost. This method uses game theory principles to maximize the aggregated users' and devices' utilities. In particular, it improves the application completion time, economic cost, service deadline, and the Edge service violation when a device fails to complete its execution within the deadline.

Nguyen et al.~\cite{nguyen2021two} proposed an optimization-based service placement method to balance the tradeoff between total operating cost and service quality. This optimization method aims to minimize the total operating cost for a service provider and maximize the quality of the service in terms of network delay.

Li et al.~\cite{li2017cost} proposed a Cloud-based video streaming service called CVS2 that enables on-demand transcoding of video streams. CVS2 focuses on the tradeoff between the processing time and cost to encode on-demand video streams with different qualities. This method aims to maintain a robust QoS for viewers and minimize the cost for the video service providers.

Ni et al.~\cite{Ni2017} proposed a priced-timed Petri net-based strategy for resource allocation in a Fog computing infrastructure. This method represents IoT tasks in the Fog as a priced-timed Petri net ordered by the tasks' completion times and the user’s costs of their execution. Priced-timed Petri nets~\cite{mayr2013priced} generalize classic Petri nets, incorporating continuous-time clocks, real-time constraints, and computation pricing. A heuristic algorithm allocates IoT applications to Fog devices to minimize the completion time and maximize financial savings for the resource and application providers.

Pham et al.~\cite{Pham2016} proposed a heuristic algorithm for task scheduling on the Cloud and Fog devices to maximize the aggregated utility of cost and completion time. This method ranks the application's tasks based on the critical path length from every task to the exit task. Moreover, the method assumes that the total cost of executing the application on the Fog devices is lower than the Cloud because the Fog provider does not charge for its local executions.

\textit{Summary.} These works optimize two weighted objectives~\cite{li2017cost,nguyen2021two,Ni2017,Pham2016,reddy2020genetic,sharghivand2020edge,zhao2016pricing} combined in a nonlinear, prediction-based or multi-objective model~\cite{demaio-kimovski2020}. However, they do not focus on the energy consumption of the application execution on the Computing Continuum, along with the application's completion time and cost.

\section{Discussion}\label{discussion}
We briefly describe the main contributions of the existing works and identify the research directions. 
We classify them based on their primary objectives for optimization and methodologies.

Table~\ref{tblc:obj} presents an overview of the prevailing research efforts discussed in the previous sections. The table classifies the research works based on the optimization 
metrics they use. Furthermore, it presents 
the factors of processing time, latency, and bandwidth influencing the completion time.

\begin{figure}[!t]
    \centering
    \includegraphics[width=.85\columnwidth]{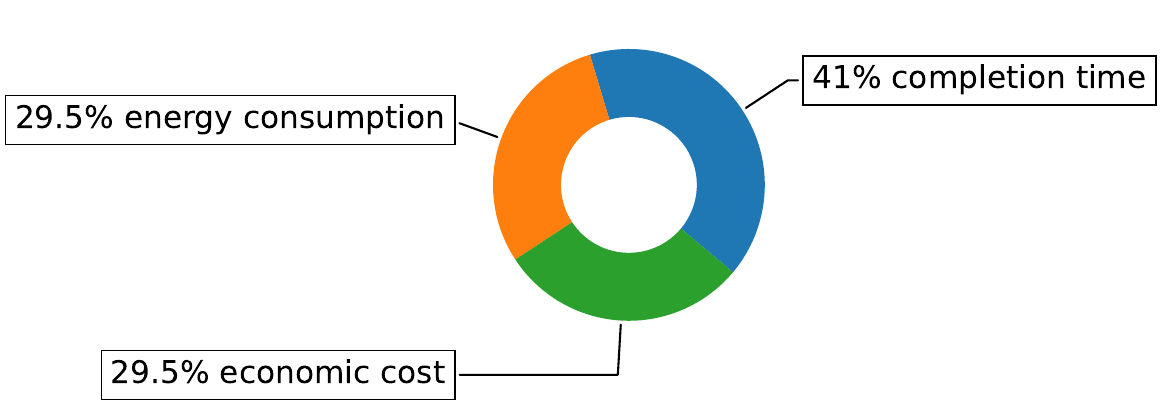}
    \caption{\centering Percentages of research articles addressing the time, energy, or cost objectives.}
    \label{fig:pie}
\end{figure}

As we observe in Table~\ref{tblc:obj}, most methods optimize the factors influencing the completion time (41\%), network latency, and bandwidth. In addition to these objectives, only a smaller fraction focuses on reducing energy consumption (29.5\%) and economic costs (29.5\%). We summarize in Figure~\ref{fig:pie} the percentage of research articles in this survey that address the aforementioned objectives. Therefore, we can conclude that the currently available techniques do not adequately address the energy efficiency issues and economic costs essential for the resource providers.

\begin{table*}[!t]
\centering
\small
\caption{Classification of the research works according to scheduling optimization metrics.}
\label{tblc:obj}
\begin{tabular}{|c||c|c|c|c|c|}
\cline{2-6}
\multicolumn{1}{c|}{}&\multicolumn{5}{c|}{\emph{\textbf{Metric}}}\\
\hline
{{\emph{\textbf{Reference}}}}& \makecell{\textbf{Completion  time}}&\textbf{Latency}&\textbf{Bandwidth}&\textbf{Energy consumption}&\textbf{Economic cost}\\
\hline\hline
Samani et al.~\cite{samani2023incremental} &\checkmark&\checkmark& \checkmark& & \\
\hline
Skarlat et al.~\cite{Skarlat2017QoS-aware} &\checkmark  &\checkmark&& &\\
\hline
Veith et al.~\cite{da2018latency} &\checkmark&\checkmark&\checkmark&&\\
\hline
Veith et al.~\cite{da2019monte} &\checkmark&\checkmark&\checkmark&&\\
\hline
Xia et al.~\cite{xia2018combining}&\checkmark&\checkmark&\checkmark&&\\
\hline
Souza et al.~\cite{souza2018towards} &\checkmark& &\checkmark&&\\
\hline
Sun et al.~\cite{Sun2018}&\checkmark&\checkmark&\checkmark&&\\
\hline
Elgamal et al.~\cite{elgamal2018droplet} &\checkmark&&\checkmark&&\\
\hline
Khare et al.~\cite{khare2019linearize}&\checkmark&\checkmark&&&\\
\hline
Ravindra et al.~\cite{ravindra2017mathbb}&\checkmark&&\checkmark&&\\
\hline
Renart et al.~\cite{renart2019distributed}&\checkmark&\checkmark&\checkmark&&\checkmark\\
\hline
Kansal et al.~\cite{kansal2010virtual}  &\checkmark&&& \checkmark&\\
\hline
Al Faruque et al.~\cite{al2015energy}&&&& \checkmark&\checkmark\\
\hline
Shojafar et al.~\cite{shojafar2016energy}&\checkmark&\checkmark&\checkmark& \checkmark&\\
\hline
Pooranian et al.~\cite{pooranian2017novel} &\checkmark&&\checkmark& \checkmark&\\
\hline
You et al.~\cite{you2018asynchronous}& &\checkmark&\checkmark&\checkmark&\checkmark\\
\hline
Reddy et al.~\cite{reddy2020genetic}&\checkmark&\checkmark&\checkmark&\checkmark&\\
\hline
Menouer~\cite{menouer2021kcss}&\checkmark&&&\checkmark&\\
\hline
Aazam et al.~\cite{aazam2015fog}&\checkmark&&\checkmark&&\checkmark\\
\hline
Zhao et al.~\cite{zhao2016pricing}&\checkmark&&&&\checkmark\\
\hline
De Maio et al.~\cite{demaio-kimovski2020} &\checkmark&\checkmark&\checkmark&&\checkmark\\
\hline
Sharghivand et al.~\cite{sharghivand2020edge} &\checkmark&&&&\checkmark\\
\hline
Nguyen et al.~\cite{nguyen2021two}&\checkmark&\checkmark&\checkmark&&\checkmark\\
\hline
Li et al.~\cite{li2017cost}&\checkmark&&&&\checkmark\\
\hline
Ni et al.~\cite{Ni2017}&\checkmark&&&&\checkmark\\
\hline
Pham et al.~\cite{Pham2016}&\checkmark&&\checkmark&&\checkmark\\
\hline
\hline
\end{tabular}
\end{table*}

\begin{figure}[!t]
    \centering
    \includegraphics[width=.85\columnwidth]{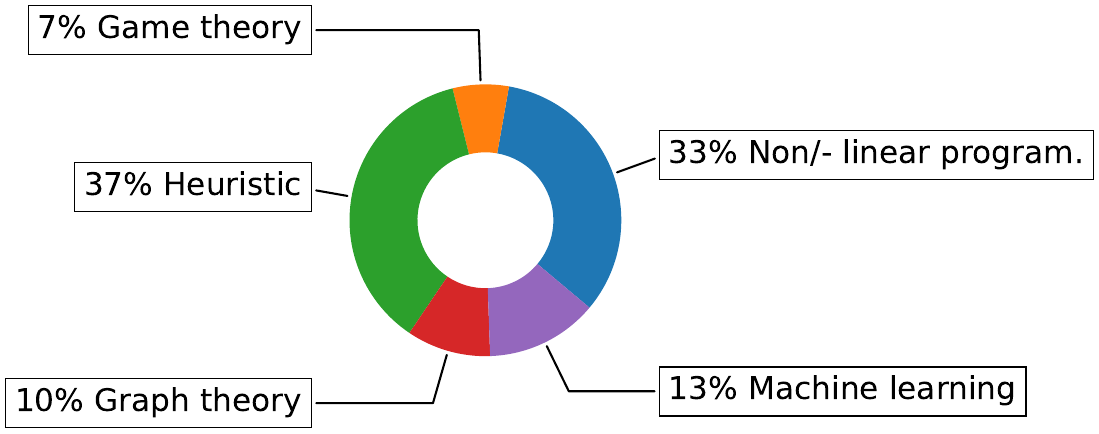}
    \caption{\centering Percentages of research articles presenting different methods.}
    \label{fig:pie2}
\end{figure}

Table~\ref{tblc:methods} further summarizes the research works based on the optimization 
methods and core algorithms used. We can observe that the majority of methods are based on heuristic techniques. However, considering the other optimization algorithms, other works use several possible methods, including graph theory and machine learning. Figure~\ref{fig:pie2} shows that most research works propose linear-, nonlinear-programming, or heuristic-based scheduling methods.

\begin{figure}[!t]
    \centering
    \includegraphics[width=.85\columnwidth]{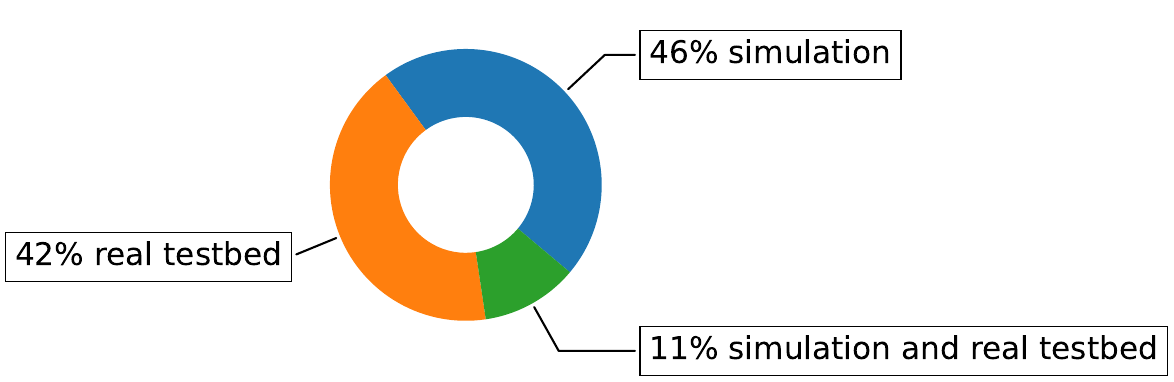}
    \caption{\centering Percentages of research articles evaluating their methods by simulation, real testbed, or both tools.}
    \label{fig:pie3}
\end{figure}

In terms of evaluation tools, Table~\ref{tblc:eval} reveals that most of the works use either simulation or a real testbed to evaluate the placement and scheduling methods. Only a small fraction implements both testbed infrastructures and simulation. Figure~\ref{fig:pie3} depicts the ratio of the research papers evaluating their methods through the simulation or real testbed experiments.

\begin{table*}[!t]
\centering
\small
\caption{Classification of the research works according to optimization methods.}
\label{tblc:methods}
\begin{tabular}{|c||c|c|c|c|c|}
\cline{2-6}
\multicolumn{1}{c|}{}&\multicolumn{5}{c|}{\emph{\textbf{Method}}}\\
\hline
{{\emph{\textbf{Reference}}}}&\makecell{\textbf{Non/- linear} \textbf{programming}}&\makecell{\textbf{Heuristic}}&\textbf{Game theory}&\textbf{Graph theory}&\textbf{Machine learning}\\
\hline\hline
Samani et al.~\cite{samani2023incremental}& && & \checkmark &\\
\hline
Skarlat et al.~\cite{Skarlat2017QoS-aware} & \checkmark & & & &\\
\hline
Veith et al.~\cite{da2018latency} & \checkmark & & &\checkmark&\\ 
\hline
Veith et al.~\cite{da2019monte} &  & & &\checkmark&\checkmark\\ 
\hline
Xia et al.~\cite{xia2018combining}& \checkmark& & & &\\
\hline
Souza et al.~\cite{souza2018towards} & & \checkmark & &  &\\
\hline
Sun et al.~\cite{Sun2018}&&\checkmark&&&\\
\hline
Elgamal et al.~\cite{elgamal2018droplet} & \checkmark & & & &\\
\hline
Khare et al.~\cite{khare2019linearize}&&\checkmark& &&\checkmark\\ 
\hline
Ravindra et al.~\cite{ravindra2017mathbb}&  &\checkmark&&&\\
\hline
Renart et al.~\cite{renart2019distributed} & \checkmark & & &  &\\
\hline
Kansal et al.~\cite{kansal2010virtual}  &  & & &  &\checkmark\\
\hline
Al Faruque et al.~\cite{al2015energy} && \checkmark& &  &\\
\hline
Shojafar et al.~\cite{shojafar2016energy}&\checkmark&& & &\\
\hline
Pooranian et al.~\cite{pooranian2017novel}  &\checkmark&& & &\\
\hline
You et al.~\cite{you2018asynchronous} &\checkmark& & &  &\\
\hline
Reddy et al.~\cite{reddy2020genetic}&&\checkmark& &  &\checkmark\\
\hline
Menouer~\cite{menouer2021kcss}&&\checkmark& &  &\\
\hline
Aazam et al.~\cite{aazam2015fog}&\checkmark&& &  &\\
\hline
Zhao et al.~\cite{zhao2016pricing}&  & &\checkmark&&\\
\hline
De Maio et al.~\cite{demaio-kimovski2020} & & \checkmark & & &\\
\hline
Sharghivand et al.~\cite{sharghivand2020edge} & & & \checkmark & &\\
\hline
Nguyen et al.~\cite{nguyen2021two}& \checkmark&  & & &\\
\hline
Li et al.~\cite{li2017cost}&&\checkmark  & & &\\\hline
Ni et al.~\cite{Ni2017}&&\checkmark  & & &\\\hline
Pham et al.~\cite{Pham2016}& &\checkmark  & & &\\
\hline
\hline
\end{tabular}
\end{table*}

\begin{table*}[!t]
\centering
\small
\caption{Classification of the research works according to evaluation tools.}
\label{tblc:eval}
\begin{tabular}{|c||c|c|c|c|c|c|c|c|c|}
\cline{2-10}
\multicolumn{1}{c|}{}&\multicolumn{9}{c|}{\emph{\textbf{Evaluation tool}}}\\
\cline{2-10}
\multicolumn{1}{c|}{}& \multicolumn{8}{c|}{\textbf{Simulation}} & \multicolumn{1}{c|}{{\textbf{Real}}}\\
\cline{1-9}
{\emph{\textbf{Reference}}}&{CloudSim}&{iFogSim}&{SimGrid}&{FogTorchPI}&{SLEIPNIR}&{OMNET++}&{YAFS}&{Matlab}&\multicolumn{1}{c|}{{\textbf{ testbed}}}\\
\hline\hline
Samani et al.~\cite{samani2023incremental}& & & & & & & \checkmark &&\checkmark\\
\hline
Skarlat et al.~\cite{Skarlat2017QoS-aware}&&\checkmark& & & & & &&\\
\hline
Veith et al.~\cite{da2018latency}&& & & & & \checkmark& &&\\ 
\hline
Veith et al.~\cite{da2019monte}&& & & & & \checkmark& &&\\ 
\hline
Xia et al.~\cite{xia2018combining}&& &\checkmark& & & & &&\\
\hline
Souza et al.~\cite{souza2018towards}& & & & &  & & & & \checkmark\\
\hline
Sun et al.~\cite{Sun2018}& && & & & &&\checkmark&\\
\hline
Elgamal et al.~\cite{elgamal2018droplet}&& & & & & & &&\checkmark\\ 
\hline
Khare et al.~\cite{khare2019linearize}&& && & & & &&\checkmark\\ 
\hline
Ravindra et al.~\cite{ravindra2017mathbb}&&& & & & & & &\checkmark\\
\hline
Renart et al.~\cite{renart2019distributed}&& & & &  & & & & \checkmark\\
\hline
Kansal et al.~\cite{kansal2010virtual}  && && & & & &&\checkmark\\
\hline 
Al Faruque et al.~\cite{al2015energy} && & && & & &&\checkmark\\
\hline 
Shojafar et al.~\cite{shojafar2016energy}&&&& & & & &\checkmark&\\
\hline 
Pooranian et al.~\cite{pooranian2017novel} &&\checkmark& & & & & &&\\
\hline 
You et al.~\cite{you2018asynchronous} & & & &&  & & & \checkmark&\\
\hline
Reddy et al.~\cite{reddy2020genetic}&&\checkmark& & & & & && \checkmark\\
\hline
Menouer~\cite{menouer2021kcss}&\checkmark&& & & & & && \checkmark\\
\hline
Aazam et al.~\cite{aazam2015fog}&\checkmark&& & & & & & &\\
\hline
Zhao et al.~\cite{zhao2016pricing}&&& & & & & & &\checkmark\\
\hline
De Maio et al.~\cite{demaio-kimovski2020} && & &\checkmark& \checkmark & & &&\\
\hline
Sharghivand et al.~\cite{sharghivand2020edge} && & & & & & && \checkmark\\
\hline
Nguyen et al.~\cite{nguyen2021two}&&&& & & & &\checkmark&\\\hline
Li et al.~\cite{li2017cost}&& & & & & & && \checkmark\\
\hline
Ni et al.~\cite{Ni2017}&& & & & & & && \checkmark\\
\hline
Pham et al.~\cite{Pham2016}&\checkmark& & & & & & && \\
\hline
\hline
\end{tabular}
\end{table*}

\section{Conclusion}\label{conclusion}
We surveyed the literature's objectives and methodologies related to resource provisioning and scheduling for dataflow processing applications on the Cloud, Fog, and Edge. We categorized the research studies based on their objectives: completion time, energy consumption, and economic cost. The classification shows that most of the research works explored the completion time with a focus on the processing time and that energy consumption or user's cost for application execution were less often addressed. We plan to extend this survey by exploring autoscaling methods for large-scale data processing in the Computing Continuum.

\section*{Acknowledgement} This work received partial funding from:
\emph{European Union}'s grant agreements H2020 101016835 (DataCloud), HE  101093202 (Graph-Massivizer), and HE 101070284 (enRichMyData);
\emph{Austrian Research Promotion Agency (FFG)}, grant agreement 888098 (K{\"a}rntner Fog) and FO999897846 (GAIA).

\bibliography{ref}

\begin{thebibliography}{10}

\bibitem{de2018distributed}
Marcos~Dias de~Assuncao, Alexandre da~Silva~Veith, and Rajkumar Buyya.
\newblock Distributed data stream processing and edge computing: A survey on resource elasticity and future directions.
\newblock {\em Journal of Network and Computer Applications}, 103:1--17, 2018.

\bibitem{mehran2022c3match}
Narges Mehran, Zahra~Najafabadi Samani, Dragi Kimovski, and Radu Prodan.
\newblock Matching-based scheduling of asynchronous data processing workflows on the computing continuum.
\newblock In {\em 2022 IEEE International Conference on Cluster Computing (CLUSTER)}, pages 58--70. IEEE, 2022.

\bibitem{exoscaleedge}
{Exoscale Corporation}.
\newblock Exoscale edge.
\newblock \url{https://www.exoscale.com/edge/}, 2023.

\bibitem{gdce}
{Google LLC}.
\newblock Extend {{Google}} cloud's infrastructure and services to the edge and your data centers.
\newblock \url{https://cloud.google.com/distributed-cloud/}, 2023.

\bibitem{kimovski2021mobility}
Dragi Kimovski, Narges Mehran, Christopher~Emanuel Kerth, and Radu Prodan.
\newblock Mobility-aware {{IoT}} applications placement in the cloud edge continuum.
\newblock {\em IEEE Transactions on Services Computing}, 15(6):3358--3371, 2022.

\bibitem{balouek2019towards}
Daniel Balouek-Thomert, Eduard~Gibert Renart, Ali~Reza Zamani, Anthony Simonet, and Manish Parashar.
\newblock Towards a computing continuum: Enabling edge-to-cloud integration for data-driven workflows.
\newblock {\em The International Journal of High Performance Computing Applications}, 33(6):1159--1174, 2019.

\bibitem{kochovski2019architecture}
Petar Kochovski, Rizos Sakellariou, Marko Bajec, Pavel Drobintsev, and Vlado Stankovski.
\newblock An architecture and stochastic method for database container placement in the edge-fog-cloud continuum.
\newblock In {\em IEEE International Parallel and Distributed Processing Symposium (IPDPS)}, pages 396--405. IEEE, 2019.

\bibitem{samani2023incremental}
Zahra~Najafabadi Samani, Narges Mehran, Dragi Kimovski, Shajulin Benedict, Nishant Saurabh, and Radu Prodan.
\newblock Incremental multilayer resource partitioning for application placement in dynamic fog.
\newblock {\em IEEE Transactions on Parallel and Distributed Systems}, 34(6):1877--1896, 2023.

\bibitem{Skarlat2017QoS-aware}
Olena Skarlat, Matteo Nardelli, Stefan Schulte, and Schahram Dustdar.
\newblock Towards {{QoS}}-aware fog service placement.
\newblock In {\em 1st International Conference on Fog and Edge Computing (ICFEC)}, pages 89--96. IEEE, 2017.

\bibitem{da2018latency}
Alexandre da~Silva~Veith, Marcos~Dias de~Assun{\c{c}}ao, and Laurent Lef{\`e}vre.
\newblock Latency-aware placement of data stream analytics on edge computing.
\newblock In {\em International Conference on Service-Oriented Computing}, pages 215--229. Springer, 2018.

\bibitem{da2019monte}
Alexandre da~Silva~Veith, Marcos~Dias de~Assun{\c{c}}ao, and Laurent Lef{\`e}vre.
\newblock {Monte-Carlo} tree search and reinforcement learning for reconfiguring data stream processing on edge computing.
\newblock In {\em 2019 31st International Symposium on Computer Architecture and High Performance Computing (SBAC-PAD)}, pages 48--55. IEEE, 2019.

\bibitem{xia2018combining}
Ye~Xia, Xavier Etchevers, Loic Letondeur, Adrien Lebre, Thierry Coupaye, and Fr{\'e}d{\'e}ric Desprez.
\newblock Combining heuristics to optimize and scale the placement of {{IoT}} applications in the fog.
\newblock In {\em 11th IEEE/ACM Conference on Utility and Cloud Computing, UCC}, pages 153--163, 2018.

\bibitem{souza2018towards}
VB~Souza, Xavier Masip-Bruin, Eva Mar{\'\i}n-Tordera, Sergio S{\`a}nchez-L{\'o}pez, Jordi Garcia, Guang-Jie Ren, Admela Jukan, and A~Juan Ferrer.
\newblock Towards a proper service placement in combined fog-to-cloud ({{F2C}}) architectures.
\newblock {\em Future Generation Computer Systems}, 87:1--15, 2018.

\bibitem{Sun2018}
Yan Sun, Fuhong Lin, and Haitao Xu.
\newblock Multi-objective optimization of resource scheduling in fog computing using an improved {{NSGA-II}}.
\newblock {\em Wireless Personal Communications}, pages 1--17, 2018.

\bibitem{elgamal2018droplet}
Tarek Elgamal, Atul Sandur, Phuong Nguyen, Klara Nahrstedt, and Gul Agha.
\newblock Droplet: Distributed operator placement for {{IoT}} applications spanning edge and cloud resources.
\newblock In {\em 2018 IEEE 11th International Conference on Cloud Computing (CLOUD)}, pages 1--8. IEEE, 2018.

\bibitem{khare2019linearize}
Shweta Khare, Hongyang Sun, Julien Gascon-Samson, Kaiwen Zhang, Aniruddha Gokhale, Yogesh Barve, Anirban Bhattacharjee, and Xenofon Koutsoukos.
\newblock Linearize, predict and place: minimizing the makespan for edge-based stream processing of directed acyclic graphs.
\newblock In {\em Proceedings of the 4th ACM/IEEE Symposium on Edge Computing}, pages 1--14, 2019.

\bibitem{ravindra2017mathbb}
Pushkara Ravindra, Aakash Khochare, Siva~Prakash Reddy, Sarthak Sharma, Prateeksha Varshney, and Yogesh Simmhan.
\newblock Echo: An adaptive orchestration platform for hybrid dataflows across cloud and edge.
\newblock In {\em International Conference on Service-Oriented Computing}, pages 395--410. Springer, 2017.

\bibitem{renart2019distributed}
Eduard~Gibert Renart, Alexandre Da~Silva Veith, Daniel Balouek-Thomert, Marcos~Dias de~Assuncao, Laurent Lef{\`e}vre, and Manish Parashar.
\newblock Distributed operator placement for {{IoT}} data analytics across edge and cloud resources.
\newblock In {\em CCGrid 2019 - 19th Annual IEEE/ACM International Symposium in Cluster, Cloud, and Grid Computing}, pages 1--10. IEEE/ACM, 2019.

\bibitem{kansal2010virtual}
Aman Kansal, Feng Zhao, Jie Liu, Nupur Kothari, and Arka~A Bhattacharya.
\newblock Virtual machine power metering and provisioning.
\newblock In {\em Proceedings of the 1st ACM Symposium on Cloud Computing}, pages 39--50, 2010.

\bibitem{al2015energy}
Mohammad~Abdullah Al~Faruque and Korosh Vatanparvar.
\newblock Energy management-as-a-service over fog computing platform.
\newblock {\em IEEE Internet of Things Journal}, 3(2):161--169, 2015.

\bibitem{shojafar2016energy}
Mohammad Shojafar, Nicola Cordeschi, and Enzo Baccarelli.
\newblock Energy-efficient adaptive resource management for real-time vehicular cloud services.
\newblock {\em IEEE Transactions on Cloud computing}, 7(1):196--209, 2016.

\bibitem{pooranian2017novel}
Zahra Pooranian, Mohammad Shojafar, Paola G~Vinueza Naranjo, Luca Chiaraviglio, and Mauro Conti.
\newblock A novel distributed fog-based networked architecture to preserve energy in fog data centers.
\newblock In {\em IEEE 14th International Conference on Mobile Ad Hoc and Sensor Systems (MASS)}, pages 604--609. IEEE, 2017.

\bibitem{you2018asynchronous}
Changsheng You, Yong Zeng, Rui Zhang, and Kaibin Huang.
\newblock Asynchronous mobile-edge computation offloading: Energy-efficient resource management.
\newblock {\em IEEE Transactions on Wireless Communications}, 17(11):7590--7605, 2018.

\bibitem{reddy2020genetic}
K~Hemant~Kumar Reddy, Ashish~Kr Luhach, Buddhadeb Pradhan, Jatindra~Kumar Dash, and Diptendu~Sinha Roy.
\newblock A genetic algorithm for energy efficient fog layer resource management in context-aware smart cities.
\newblock {\em Sustainable Cities and Society}, 63:102428, 2020.

\bibitem{menouer2021kcss}
Tarek Menouer.
\newblock {KCSS}: Kubernetes container scheduling strategy.
\newblock {\em The Journal of Supercomputing}, 77(5):4267--4293, 2021.

\bibitem{hwang2012multiple}
C-L Hwang and Abu Syed~Md Masud.
\newblock {\em Multiple objective decision making—methods and applications: a state-of-the-art survey}, volume 164.
\newblock Springer Science \& Business Media, 2012.

\bibitem{aazam2015fog}
Mohammad Aazam and Eui-Nam Huh.
\newblock Fog computing micro datacenter based dynamic resource estimation and pricing model for {{IoT}}.
\newblock In {\em IEEE 29th International Conference on Advanced Information Networking and Applications}, pages 687--694. IEEE, 2015.

\bibitem{zhao2016pricing}
Tianchu Zhao, Sheng Zhou, Xueying Guo, Yun Zhao, and Zhisheng Niu.
\newblock Pricing policy and computational resource provisioning for delay-aware mobile edge computing.
\newblock In {\em 2016 IEEE/CIC International Conference on Communications in China (ICCC)}, pages 1--6. IEEE, 2016.

\bibitem{demaio-kimovski2020}
Vincenzo {De Maio} and Dragi Kimovski.
\newblock Multi-objective scheduling of extreme data scientific workflows in fog.
\newblock {\em Future Generation Computer Systems}, 106:171 -- 184, 2020.

\bibitem{Brogi2017}
A.~Brogi, S.~Forti, and A.~Ibrahim.
\newblock How to best deploy your fog applications, probably.
\newblock In {\em 1st IEEE International Conference on Fog and Edge Computing (ICFEC)}, pages 105--114, 2017.

\bibitem{sharghivand2020edge}
Nafiseh Sharghivand, Farnaz Derakhshan, Lena Mashayekhy, and Leyli Mohammadkhanli.
\newblock An edge computing matching framework with guaranteed quality of service.
\newblock {\em IEEE Transactions on Cloud Computing}, 10(3):1557--1570, 2022.

\bibitem{nguyen2021two}
Duong~Tung Nguyen, Hieu~Trung Nguyen, Ni~Trieu, and Vijay~K Bhargava.
\newblock Two-stage robust edge service placement and sizing under demand uncertainty.
\newblock {\em IEEE Internet of Things Journal}, 9(2):1560--1574, 2021.

\bibitem{li2017cost}
Xiangbo Li, Mohsen~Amini Salehi, Magdy Bayoumi, Nian-Feng Tzeng, and Rajkumar Buyya.
\newblock Cost-efficient and robust on-demand video transcoding using heterogeneous cloud services.
\newblock {\em IEEE Transactions on Parallel and Distributed Systems}, 29(3):556--571, 2017.

\bibitem{Ni2017}
Lina Ni, Jinquan Zhang, Changjun Jiang, Chungang Yan, and Kan Yu.
\newblock Resource allocation strategy in fog computing based on priced timed {Petri} nets.
\newblock {\em IEEE Internet of Things Journal}, 4(5):1216--1228, 2017.

\bibitem{mayr2013priced}
Richard~M Mayr and Parosh~Aziz Abdulla.
\newblock Priced timed {Petri} nets.
\newblock {\em Logical Methods in Computer Science}, 9, 2013.

\bibitem{Pham2016}
Xuan-Qui Pham and Eui-Nam Huh.
\newblock Towards task scheduling in a cloud-fog computing system.
\newblock In {\em Network Operations and Management Symposium (APNOMS), 2016 18th Asia-Pacific}, pages 1--4. IEEE, 2016.

\end{thebibliography}
\bibliographystyle{unsrt}
\end{document}